\documentclass[prd,aps,noshowpacs,nofootinbib,11pt]{revtex4}
\usepackage{amsmath,graphicx,color,epsfig}

\begin{document}

\title{Lepton flavor violating processes in unparticle physics}
\author{Cai-Dian L\"u$^{a}$,
Wei Wang$^{a,b}$ and Yu-Ming Wang$^{a,b}$ } \affiliation{\it \small $^a$
 Institute of High Energy Physics, P.O. Box 918(4) Beijing, 100049, P.R.
 China\\
 \it \small $^b$  Graduate University of Chinese Academy of Sciences, Beijing,
 100049, P.R. China}

\begin{abstract}
We study the virtual effects of unparticle physics in the  lepton
flavor violating processes $M^0\to l^+l'^-$ and $e^+e^-\to l^+l'^-$
scattering, where $M^0$ denotes the pseudoscalar mesons: $\pi^0,K_L,
D_0,B_0,B_s^0$ and $l,l'$ denote two different lepton flavors. For
the decay of $B^0\to l^+l'^-$, there is no constraint from the
current experimental upper bounds on the vector unparticle coupling
with leptons. The constraint on the coupling constant between scalar
unparticle field and leptons is sensitive to the scaling dimension
of the unparticle $d_{\cal U}$. For the scattering process
$e^-e^+\to l^-l'^+$, there is only constraint from experiments on
the vector unparticle couplings with leptons but no constraint on
the scalar unparticle. We study the $\sqrt s$ dependence of the
cross section $ \frac{1}{\sigma} \frac{d\sigma}{d\sqrt s}$ of
$e^+e^-\to l^-l'^+$ with different values of $d_{\cal U}$. If
$d_{\cal U}=1.5$, the cross section is independent on the center
mass energy. For $d_{\cal U}>1.5$, the cross section increases with
$\sqrt s$.

\end{abstract}
\maketitle

\section{Introduction}

In four space-time dimensions, there is no scale invariant
interacting quantum field theory which contains massive particles.
Even if scale invariance is preserved in massless field theory at
classical level, it would be broken by renormalization effect which
is known as the trace anomaly. Nevertheless, it is possible that at
a much higher scale there exists a scale invariant sector with a
nontrivial infrared fixed point. Recently, it has been argued that
one kind of fields, under the name Banks-Zaks ($\mathcal{BZ}$)
fields \cite{Banks:1981nn}, might appear at TeV scale. At low
energy, these fields manifest themselves by matching onto a new
sector called ``unparticle'' (${\cal U}$) with a non-integral number
of scale dimension $d_{\cal U}$ \cite{Georgi:2007ek,Georgi:2007si}.

Although the underlying structure of unparticles is still unclear,
 there indeed exist respectable interesting phenomena for testing
unparticles experimentally. In many processes, $t\to u{\cal U}$
\cite{Georgi:2007ek}, $e^+e^-\to \gamma {\cal U}$ and $Z\to \bar
qq {\cal U}$ \cite{Cheung:2007ue}, the productions of these stuff
might be detected by measuring the missing energies and the
momentum distributions. The phenomenological studies on the
unparticle effect in charged Higgs decays, anomalous magnetic
moments, $B^0-\bar B^0$ and $D^0-\bar D^0$ mixing and hadronic
flavor changing neutral current (FCNC) in $B$ decays are carried
out in
Ref.~\cite{Cheung:2007ue,Luo:2007bq,Chen:2007vv,Ding:2007bm,Liao:2007bx,Li:2007by}.
In Ref.~\cite{Aliev:2007qw}, the lepton flavor violation
interaction is introduced   to explore the phenomenology in
$\mu^+\to e^-e^+e^-$. In this work, we will investigate the lepton
flavor changing processes $B^0\to\mu^\mp\tau^\pm$, $e^+e^-\to
e^\mp\mu^\pm$ and some other related processes. There are a number
of experimental upper bounds on these processes which give
stringent constraints to the effective couplings of unparticles.
This can probably shed light on the internal structure of
unparticles.

\section{Effective interactions}

At very high energy, the theory contains the Standard Model (SM)
fields and the fields with a nontrivial infrared fixed point, called
$\mathcal{BZ}$ fields \cite{Banks:1981nn}. These two sectors
interact with each other by exchanging particles with a very large
mass $M_{\cal U}$.  Below this mass scale, heavy particles are
integrated out and thus the SM particles  and the $\mathcal{BZ}$
sector interacts through non-renormalizable operators:
\begin{eqnarray}
\frac{1}{M^{d_{SM}+d_{\cal BZ}-4}_{\cal U}}O_{SM} O_{{\cal
BZ}},\label{Eq:nonren}
\end{eqnarray}
where $O_{SM}$ and $O_{{\cal BZ}}$ are local operators made up of
the SM and $\mathcal{BZ}$ fields, respectively. The
renormalization effects in the scale invariant sector induce the
dimensional transmutation at the scale $\Lambda_{{\cal U}}$. Below
the scale $\Lambda_{\cal U}$, the $\mathcal{BZ}$ operators match
onto unparticle operators while the non-renormalizable operators
in Eq.~(\ref{Eq:nonren}) match onto an effective interaction
operator:
\begin{eqnarray}
\frac{C_{{\cal U}}\Lambda_{{\cal U}}^{d_{{\cal BZ}}-d_{{\cal
U}}}}{M_{{\cal U}}^{d_{SM}+d_{\cal BZ}-4}}O_{SM} O_{{\cal
U}},\label{Eq:Leff1}
\end{eqnarray}
where $d_{\mathcal{BZ}}$ and $d_{\cal U}$ are the scaling
dimensions of the $O_{\mathcal{BZ}}$ and unparticle $O_{\cal U}$
operators respectively. The scaling dimension of unparticle field
has been taken as $1<d_{\cal U}<2$ in the literature.

To be more specific, unparticles have some different characters
from ordinary particles in phase space, virtual propagator, and
the effective interaction with the SM particles.  It was
demonstrated in \cite{Georgi:2007ek} that scale invariance can be
used to fix the two-point functions of the unparticle operators
and the propagators. The propagator of the scalar unparticle field
can be written by \cite{Georgi:2007si,Cheung:2007ue}:
\begin{eqnarray}
\int e^{iP\cdot x}d^4x\langle 0 |T[O_{\cal U}(x) O_{\cal U}(0)
 ]|0\rangle =i\frac{A_{d_{\cal U}}}{2}\frac{1}{\mbox{sin}(d_{\cal U}\pi)}(-P^2-i\epsilon)^{d_{\cal
 U}-2}.
\end{eqnarray}
If  the vector unparticle  is assumed to be transverse, its
propagator has the form
\begin{eqnarray}
\int e^{iP\cdot x}d^4x\langle 0 |T[O^\mu_{\cal U}(x) O^\nu_{\cal
U}(0)
 ]|0\rangle =
i\frac{A_{d_{\cal U}}}{2}\frac{-g^{\mu\nu}+P^\mu
P^\nu/P^2}{\mbox{sin}(d_{\cal U}\pi)}(-P^2-i\epsilon)^{d_{\cal
U}-2},
\end{eqnarray}
where the coefficient $A_{d_{\cal U}}$ is given by
\begin{eqnarray}
A_{d_{\cal U}}=\frac{16\pi^{5/2}}{(2\pi)^{2d_{\cal
U}}}\frac{\Gamma(d_{\cal U}+\frac{1}{2})}{\Gamma(d_{\cal
U}-1)\Gamma(2d_{\cal U})}.
\end{eqnarray}
There are other possible Lorentz structures: a spinor field
\cite{Luo:2007bq} or even a tensor field $O_{\cal U}^{\mu\nu}$.
The only difference is the spin structure which has been
comprehensively discussed in Ref. \cite{Cheung:2007ap}.

The effective interactions that satisfy the standard model gauge
symmetry for the scalar and vector unparticle operators with
standard model fields are given, respectively, by
\begin{eqnarray}
&& \lambda_0 \frac{ 1}{\Lambda_{\cal U}^{d_{\cal U}-1}} \bar f f
O_{\cal U}\;, \;\; \lambda_0 \frac{1}{\Lambda_{\cal U}^{d_{\cal
U}-1} } \bar f i
      \gamma_5 f O_{\cal U}\;, \;\;
\lambda_0 \frac{1}{\Lambda_{\cal U}^{d_{\cal U}} } \bar f
\gamma^\mu f (\partial_\mu O_{\cal U}) \;, \lambda_0
\frac{1}{\Lambda_{\cal U}^{d_{\cal U}} } \bar f \gamma^\mu\gamma_5
f (\partial_\mu O_{\cal U}) \;,
\nonumber \\
\ \ \ &&\lambda_0 \frac{1}{\Lambda_{\cal U}^{d_{\cal U}} }
G_{\alpha\beta} G^{\alpha\beta} O_{\cal U} \;, \lambda_1
\frac{1}{\Lambda_{\cal U}^{d_{\cal U} - 1} }\, \bar f \gamma_\mu f
\, O_{\cal U}^\mu \;, \;\; \lambda_1 \frac{1}{\Lambda_{\cal
U}^{d_{\cal U} - 1} }\, \bar f \gamma_\mu \gamma_5 f \, O_{\cal
U}^\mu \;,
\end{eqnarray}
where $f$ stands for a standard model fermion and $\lambda_{i}$ are
dimensionless effective couplings $C_{{\cal U}} \Lambda_{\cal
U}^{d_{\cal BZ}}/M_{\cal U}^{d_{SM} + d_{\cal BZ}-4}$ with the index
$i=0,1$ labeling the scalar and vector unparticle operators,
respectively.  In principle, the coupling constants
$\lambda_0,\lambda_1$ can be different for different flavors and are
then distinguished by additional indices. For example, the hadronic
FCNC via scalar unparticle, taking $b\to d$ as an example, can
proceed through the effective interaction term:
\begin{eqnarray}
L_{eff}=i\frac{\lambda_{db}}{\Lambda_{\cal U}^{d_{\cal U}}}\bar
d\gamma_\mu(1-\gamma_5)b \partial ^\mu O_{\cal
U}+h.c.,\label{eq:scalar}
\end{eqnarray}
where we have used the subscript $db$ to denote the coupling with
$d$ and $b$ quark.  Similarly, for the vector unparticle, the
effective interaction is considered as,

\begin{eqnarray}
L_{eff}=\frac{1}{\Lambda_{\cal U}^{d_{\cal U}}}\bar l
\gamma_\mu(\lambda_{Vll'}+\lambda_{All'}\gamma_5)l' ~  O^\mu_{\cal
U}+h.c. .\label{eq:vector}
\end{eqnarray}

\section{ Lepton flavor violation decays $M^0\to  l^+l'^-$}

In this section, we will consider the lepton flavor violation in
neutral meson decays, including $\pi^0$, $K_L$, $D^0$, $B_0$ and
$B_s^0$. To be more specific, we will consider the scalar coupling
in Eq.(\ref{eq:scalar}) and the vector coupling in
Eq.(\ref{eq:vector}). As discussed above, the vector unparticle is
assumed to be transverse. Thus this kind of unparticle gives zero
contribution, when contraction with $p_M^\mu=p_{\cal U}^\mu$ from
the matrix element $\langle 0|\bar q_1\gamma_\mu\gamma_5
q_2|M^0\rangle$.

In the following, we will first focus on the decay channel $B^0\to
\mu^+ \tau^-$ induced by scalar unparticle. The decay amplitude for
$B^0\to \mu^\pm \tau^\mp$ reads
\begin{eqnarray}
 i{\cal M}&=&\bar u(p_{\tau})[i\frac{\lambda_{db}}{\Lambda_{\cal U}^{d_{\cal  U}}}
             \gamma_\mu(1-\gamma_5)P^\mu_B ]v(p_\mu) \nonumber\\
           && \times \frac{iA_{d_{\cal  U}}}{2}\times \frac{1}{\mbox{sin}(d_{\cal  U}\pi)}
            \times (-m_B^2-i\epsilon)^{d_{\cal  U}-2}
           \times i\frac{\lambda_{\mu\tau}}{\Lambda_{\cal U}^{d_{\cal
           U}}}(-P^\mu_B)(-if_B {P_{B}}_{\mu})\nonumber\\
           &=&-m_{\tau}\frac{A_{d_{\cal U}}}{2\mbox{sin}(d_{\cal U}\pi)}\frac{\lambda_{db}\lambda_{\mu\tau}}{\Lambda_{\cal U}^{2d_{\cal U}}}
           f_B m_B^2(-m_B^2-i\epsilon)^{d_{\cal U}-2}
           \times\bar u(p_{\tau})(1+\gamma_5)v(p_\mu),
\end{eqnarray}
where the mass of lighter lepton $\mu$ is neglected. Thus, the decay
width for this decay channel can be written as
\begin{eqnarray}
 \Gamma&=&\frac{|\vec p|}{8\pi m_B^2}\times \sum_{pol.}|{\cal M}|^2
       =\frac{f_B^2 m_B^5 m_\tau^2 }{4\pi}|\frac{A_{d_{\cal U}}}{2\mbox{sin}(d_{\cal
U}\pi)}\frac{\lambda_{db}\lambda_{\mu\tau}}{\Lambda_{\cal
U}^{2d_{\cal U}}}
            (-m_B^2)^{d_{\cal U}-2}|^2.
\end{eqnarray}
This decay width is proportional to the lepton mass square, due to
the $V-A$ current for the interaction type, which indicates this
kind of process is helicity-suppressed.

As an illustration, we use the following inputs
\begin{eqnarray}
\Lambda^{d_{\cal  U}}_{\cal U}=1 {\mbox {TeV}}, \;\;\; d_{\cal
U}=0.5.
\end{eqnarray}
The  experimental upper  bound on the branching fraction:
\begin{eqnarray}
{\cal BR}(B^0\to \mu^{\pm}\tau^{\mp})<3.8\times 10^{-5},
\end{eqnarray}
leads to a constraint on the coupling constant
$|\lambda_{db}\lambda_{\mu\tau}| \leq 1.4 \times 10 ^{-5}$. If we
take the scaling dimension $d_{\cal U}$ larger, the constraint
becomes less stringent, for example, the constraint for
$|\lambda_{db}\lambda_{\mu\tau}| \leq 20$ is obtained by taking
$d_{\cal U}=1.5$. As mentioned above, the $V-A$ coupling of
unparticles with the standard model fermions results in the famous
helicity suppression. If other interactions are introduced such as
$S \pm P$, the helicity rule is invalid, which can constrain
$\lambda$ more strictly.

The analysis can be easily generalized to other processes, such as
$\pi^0(K_L,D^0,B^0,B_s^0)\to l^+l'^-$. These processes can give
different constraints. For example, the constraint on
$|(\lambda_{uu}-\lambda_{dd})\lambda_{e\mu}|$ is obtained from
$\pi^0\to l^+l'^-$. In table \ref{Tab:Mtoll'}, we collect the
experimental upper bounds \cite{Yao:2006px} for these channels and
their constraints   on the leptonic flavor violating processes in
$\pi^0$, $K_L$, $D^0$, $B^0$, $B_s^0$ decays. From this table, we
can see the results dramatically depend on the scaling dimension of
the unparticle field.

Since the transversely polarized vector unparticle   gives zero
contribution to the neutral meson decays, there is no constraint
from experiments to their effective couplings.

\begin{table}[tb]
\caption{Experimental upper bounds on $M\to l^+l'^-$ at $90\%$
confidence level \cite{Yao:2006px}  and their constraints on the
effective coupling constants $\lambda$   performed at
$\Lambda_{\cal U}=1$ TeV and $d_{\cal U}=0.5$, $1.5$.}
 \label{Tab:Mtoll'}
\begin{center}
 \begin{tabular}{c|c|c|c}
  \hline\hline
        {Modes}     & Experiments &   $\lambda$ ($d_{\cal U}=0.5$) &  $\lambda$ ($d_{\cal U}=1.5$)
          \\  \hline
   $\pi^0\to \mu^+ e^-$
                                                          &$3.8\times 10^{-10}$
                                                          &$|(\lambda_{uu}-\lambda_{dd})\lambda_{e\mu}|<3.7\times 10^{-5}$
                                                          &$|(\lambda_{uu}-\lambda_{dd})\lambda_{e\mu}|<2.4\times 10^{6}$
                                                          \\\hline
   $\pi^0\to \mu^- e^+$
                                                          & $3.4\times 10^{-9}$
                                                          & $|(\lambda_{uu}-\lambda_{dd})\lambda_{e\mu}|<1.1\times 10^{-4}$
                                                          & $|(\lambda_{uu}-\lambda_{dd})\lambda_{e\mu}|<7.3\times 10^{6}$
                                                          \\ \hline
  $K_L\to e^{\pm}\mu^{\mp}$
                                                          & $4.7\times 10^{-12}$
                                                          &  Re$(\lambda_{sd})|\lambda_{e\mu}|<1.3\times 10^{-10}$
                                                          & Re$(\lambda_{sd})|\lambda_{e\mu}|<0.02$
                                                          \\\hline
  $D^0\to e^{\pm}\mu^{\mp}$
                                                          & $8.1\times 10^{-7}$
                                                          & $|\lambda_{cu}\lambda_{e\mu}|<3.6\times 10^{-5}$
                                                          & $|\lambda_{cu}\lambda_{e\mu}|<409$
                                                          \\\hline
  $B^0\to e^{\pm}\mu^{\mp}$
                                                          & $4.0\times 10^{-6}$
                                                          &
                                                          $|\lambda_{db}\lambda_{e\mu}|<7.6\times 10^{-5}$
                                                          & $|\lambda_{db}\lambda_{e\mu}|<108$
                                                          \\\hline
  $B^0\to e^{\pm}\tau^{\mp}$
                                                          & $1.1\times 10^{-4}$
                                                          & $|\lambda_{db}\lambda_{e\tau}|<2.4\times 10^{-5}$
                                                          & $|\lambda_{db}\lambda_{e\tau}|<34$
                                                          \\\hline
  $B^0\to \mu^{\pm}\tau^{\mp}$
                                                          & $3.8\times 10^{-5}$
                                                          & $|\lambda_{db}\lambda_{\mu\tau}|<1.4\times 10^{-5}$
                                                          & $|\lambda_{db}\lambda_{\mu\tau}|<20$
                                                          \\\hline
  $B_s^0\to e^{\pm}\mu^{\mp}$
                                                          & $6.1\times 10^{-6}$
                                                          & $|\lambda_{sb}\lambda_{e\mu}|<7.8\times 10^{-5}$
                                                          & $|\lambda_{sb}\lambda_{e\mu}|<107$
                                                          \\
 \hline\hline\end{tabular}
\end{center}
 \end{table}

\section{ Lepton flavor violating scattering process $e^+e^-\to l^+l'^-$}

 In the following, we will consider the
process $e^-(p_1)e^+(p_2)\to e^-(p_3)\mu^+(p_4)$ as an example. In
the calculations, we will neglect the small masses of the leptons
\footnote{The mass of $\tau$ is also negligible compared with the
large center mass energy.}. The coupling between scalar unparticle
and SM particles in Eq.(\ref{eq:scalar}) is proportional to the
momentum of unparticle, when contraction with the Dirac matrix and
using equation of motion, the amplitude is proportional to the mass
of the lepton and thus negligible. This kind of helicity suppression
bring on the negligible contributions of the scalar unparticle.
Therefore, only the vector unparticle coupling given in
Eq.~(\ref{eq:vector}) will be considered in this scattering process.

\begin{figure}[htb]
\begin{center}
\vspace{-2.cm} \psfig{file=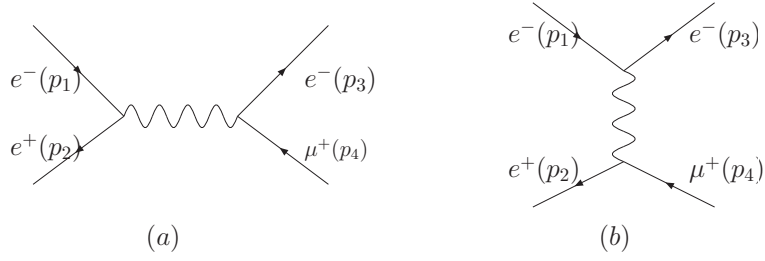,width=15.0cm,angle=0}
\end{center}
\vspace{-16.5cm} \caption{Lowest order Feynman diagrams of
$e^+e^-\to e^+\mu^-$ }\label{Feyn:eetoemu}
\end{figure}

\begin{figure}[tb]
\begin{center}
\vspace{-1.5cm}
\begin{tabular}{ccc}
\includegraphics[scale=0.6]{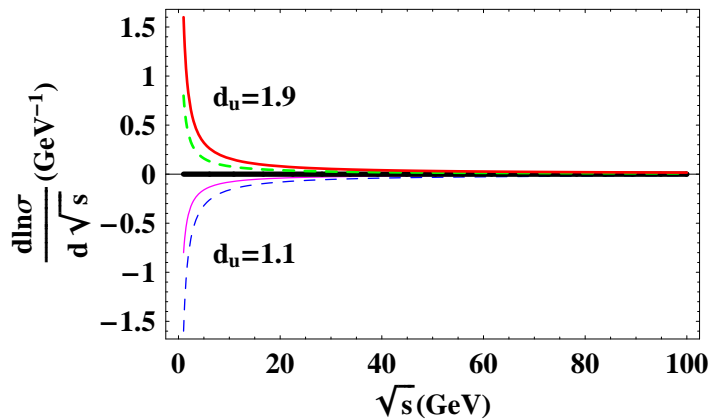}
\end{tabular}
\vspace{-1.5cm}
\caption{$\sqrt{s}$ dependence of $d {\rm{ln}}
\sigma / d\sqrt{s}$ with various values of $d_{\cal U}=1.1, 1.3,
1.5, 1.7, 1.9$ (from bottom to top), respectively
}\label{dia:sdependence}
\end{center}
\end{figure}

There are two leading order Feynman diagrams contributing to this
process, which are depicted in Fig.~\ref{Feyn:eetoemu}. With the
interaction Lagrangian in Eq.~(\ref{eq:vector}), the amplitude for
the left diagram is
\begin{eqnarray}
 i{\cal M}_a&=& \bar u(p_3)\Big[\frac{\lambda_{Ve\mu}}{\Lambda^{d_{\cal U}}}\gamma_\mu
          +\frac{\lambda_{Ae\mu}}{\Lambda^{d_{\cal U}}}\gamma_\mu\gamma_5 ]v(p_4)\nonumber\\
            &&\times \frac{iA_{d_{\cal U}}}{2}\times \frac{-g^{\mu\nu}+P^\mu
            P^\nu/P^2} {{\mbox {sin}}(d_{\cal U}\pi)}\times[-(p_1+p_2)^2-i\epsilon]^{d_{\cal U}-2}\nonumber\\
            &&\times \bar v(p_2)[\frac{\lambda_{Vee}}{\Lambda^{d_{\cal U}}}\gamma_\mu
         +\frac{\lambda_{Aee}}{\Lambda^{d_{\cal U}}}\gamma_\mu\gamma_5
         ]u(p_1),
\end{eqnarray}
while the amplitude for the right diagram is
\begin{eqnarray}
 i{\cal M}_b&=& -\bar u(p_3)[\frac{\lambda_{Vee}}{\Lambda^{d_{\cal U}}}\gamma_\mu
         +\frac{\lambda_{Aee}}{\Lambda^{d_{\cal U}}}\gamma_\mu\gamma_5]u(p_1)\nonumber\\
            &&\times \frac{iA_{d_{\cal U}}}{2}\times \frac{-g^{\mu\nu}+P^\mu
            P^\nu/P^2} {{\mbox {sin}}(d_{\cal U}\pi)}\times[-(p_3-p_1)^2-i\epsilon]^{d_{\cal U}-2}\nonumber\\
            &&\times \bar v(p_2)[\frac{\lambda_{Ve\mu}}{\Lambda^{d_{\cal U}}}\gamma_\mu
          +\frac{\lambda_{Ae\mu}}{\Lambda^{d_{\cal U}}}\gamma_\mu\gamma_5
         ]v(p_4).
\end{eqnarray}
The second term in the vector unparticle propagator will give a
contribution which is proportional to the lepton mass and thus will
be neglected in our calculation. Since there are only three
independent four-vectors in this kind of scattering, there is no
interference between the vector and the axial-vector couplings. Thus
we can consider these two different contributions independently. For
simplicity, we only consider the vector coupling by taking
$\lambda_{All'}=0$. The matrix element squared is then written by
\begin{eqnarray}
\sum_{pol.}|{\cal M}|^2=\sum_{pol.}|i{\cal
M}_a|^2+\sum_{pol.}|i{\cal M}_b|^2+\sum_{pol.}{\cal M}_a{\cal
M}_b^*+\sum_{pol.}{\cal M}_a^*{\cal M}_b,\label{eq:amplitude}
\end{eqnarray}
with
\begin{eqnarray}
 \sum_{pol.}|i{\cal M}_a|^2&=& 32|b|^2|(-s)^{d_{\cal
 U}-2}|^2s^2(1+\cos\theta)
    ,\\
   \sum_{pol.}|i{\cal M}_b|^2&=& 32|b|^2|[-(p_3-p_1)^2]^{d_{\cal U}-2}|^2s^2(1+\mbox{cos}\theta)
   ,\\
     \sum_{pol.}{\cal M}_a{\cal M}_b^*&=&32|b|^2(-s-i\epsilon)^{d_{\cal U}-2}[-(p_3-p_1)^2+i\epsilon]^{d_{\cal
     U}-2}s^2(1+\mbox{cos}\theta)\mbox{cos}(d_{\cal U}\pi)
     ,\\
     \sum_{pol.}{\cal M}_a^*{\cal M}_b&=&32|b|^2(-s+i\epsilon)^{d_{\cal U}-2}[-(p_3-p_1)^2-i\epsilon]^{d_{\cal
     U}-2}s^2(1+\mbox{cos}\theta)\mbox{cos}(d_{\cal U}\pi)
     ,
\end{eqnarray}
where $\theta$ is the scattering angle (the angle between the
3-momentum of electron: $\vec p_1$ and $\vec p_3$) and
\begin{eqnarray}
 b&=&\frac{-iA_{d_{\cal
U}}}{2}\frac{\lambda_{Vee}\lambda_{Ve\mu}}{\Lambda_{\cal
U}^{2d_{\cal U}-2}}
                \frac{1} {{\mbox {sin}}(d_{\cal U}\pi)}.
                \end{eqnarray}
The cross section is
\begin{eqnarray}
\sigma&=&\frac{1}{2s}\int
\frac{d^3\vec p_3}{(2\pi)^32E_3}\frac{d^3\vec p_4}{(2\pi)^32E_4}(2\pi)^4\delta^4(p_1+p_2-p_3-p_4)\sum_{pol.}|{\cal M}|^2\nonumber\\
 &=&\frac{1}{2s}\int
\frac{d{\rm{cos}}\theta}{16\pi}\sum_{pol.}|{\cal M}|^2.
\end{eqnarray}
One interesting thing is the dependence on the invariant mass with
different scaling dimension $d_{\cal U}$. In order to show the
dependence on $d_{\cal U}$ of the cross section, we plot the
$\sqrt s$ dependence of the function $R(s)\equiv \frac{1}{\sigma}
\frac{d\sigma}{d\sqrt s}$ of $e^+e^-\to e^-\mu^+$ with different
values of $d_{\cal U}$ in Fig.~\ref{dia:sdependence}.  The
amplitude square in Eq.(\ref{eq:amplitude}) has the behavior of
$s^{2d_{\cal U}-2}$ which gives $\sigma\sim s^{2d_{\cal U}-3}$. If
$d_{\cal U}=1.5$, the cross section is independent on the
invariant mass. For $d_{\cal U}>1.5$, the cross section increases
with $\sqrt s$. At low energy, the function $R$ has a strong
dependence on the scaling dimension $d_{\cal U}$ while at high
energy it is close to 0 and almost independent on $d_{\cal U}$ .

\begin{table}[tb]
\caption{Experimental upper bounds for cross section of $e^+e^-\to
l^+l'^-$ (in units of fb) \cite{Aubert:2006uy,Abbiendi:2001cs}. }
 \label{Tab:eetoll'}
\begin{center}
 \begin{tabular}{c|c|c|c}
  \hline\hline
          & $e\mu$ &   $e\tau$ &$\mu\tau$
          \\  \hline
   $\sqrt s=10.58 {\rm GeV}$
                                                          &---
                                                          &$9.2$
                                                          & $3.8$
                                                          \\\hline
  $\sqrt s=189 {\rm GeV}$
                                                          & $58$
                                                          & $95$
                                                          & $115$
                                                          \\ \hline
  $192 {\rm GeV} <\sqrt s<196 {\rm GeV}$
                                                          & $62$
                                                          & $144$
                                                          & $116$
                                                           \\\hline
    $200 {\rm GeV} <\sqrt s<206 {\rm GeV}$
                                                          & $22  $
                                                          & $78$
                                                          & $64$
                                                          \\
 \hline\hline\end{tabular}
\end{center}
 \end{table}

On the experimental side, OPAL and BABAR collaborations have
performed  some studies on the lepton flavor changing processes at
different invariant masses \cite{Aubert:2006uy,Abbiendi:2001cs}. The
upper bounds are collected in table~\ref{Tab:eetoll'}. We take
\begin{eqnarray}
\Lambda^{d_{\cal  U}}_{\cal U}=1 {\mbox {TeV}},
\end{eqnarray}
to give the combined constraint on the coupling constant as in
table \ref{Tab:constrainteetoll'}. The results in the table
strongly depend on the scaling dimension $d_{\cal U}$.

\begin{table}[tb]
\caption{Constraints on the lepton coupling constants from
$e^+e^-\to l^+l'^-$.}
 \label{Tab:constrainteetoll'}
\begin{center}
 \begin{tabular}{ccccccc}
  \hline
  $d_{\cal U}$             &   $1.1$               &$1.3$  &$1.5$   &$1.7$   &$1.9$\\  \hline
  $\sqrt{|\lambda_{Vee}\lambda_{Ve\mu}|}$&$<0.007$ &$<0.03$&$<0.28$ &$<0.37$ &$<0.53$ \\\hline
\end{tabular}
\end{center}
\end{table}

The constraint on the axial-vector coupling constants can also be
similarly analyzed. Since the scalar unparticle coupling does not
contribute to the $e^+e^-$ process in the zero lepton mass limit,
there is no constraint for their effective coupling from these
experiments.

\section{Conclusion}

 In a short summary, we have explored the
phenomenology of unparticle physics with the help of lepton flavor
changing processes $M^0\to l^+ l'^-$, $e^+e^-\to e^\pm\mu^\mp$ and
other related processes. In the zero lepton mass limit, the vector
unparticle coupling does not contribute  to the neutral meson
leptonic decays; while the scalar unparticle coupling does not
contribute to the $e^+e^-$ scattering processes. Therefore, the
experimental upper bounds of $M^0\to l^+ l'^-$ decays will only
constrain the scalar unparticle coupling  $\lambda$, which is
sensitive to the scaling dimension of the unparticle $d_{\cal U}$.
For the scattering process $e^-e^+\to e^-\mu^+$, there is only
constraint to the vector unparticle coupling from current
experiments. We also study the $\sqrt s$ dependence of the cross
section $ \frac{1}{\sigma} \frac{d\sigma}{d\sqrt s}$ of process
$e^-e^+\to e^-\mu^+$
 with different values of $d_{\cal U}$. If $d_{\cal U}=1.5$, the cross section
is independent on the invariant mass. For $d_{\cal U}>1.5$, the
cross section increases with $\sqrt s$.

\section*{Acknowledgements}

This work is partly supported by National Science Foundation of
China under Grant No.10475085 and 10625525. The authors would like
to thank  Ying Li, Yi Liao, Yue-Long Shen and Shun Zhou for
helpful discussions.




\begin{thebibliography}{99}
\bibitem{Banks:1981nn}
  T.~Banks and A.~Zaks,
  Nucl.\ Phys.\  B {\bf 196}, 189 (1982).

\bibitem{Georgi:2007ek}
  H.~Georgi,
  Phys.\ Rev.\ Lett.\  {\bf 98}, 221601 (2007)
  [arXiv:hep-ph/0703260].



\bibitem{Georgi:2007si}
  H.~Georgi,
  arXiv:0704.2457 [hep-ph].


\bibitem{Cheung:2007ue}
  K.~Cheung, W.~Y.~Keung and T.~C.~Yuan,
  arXiv:0704.2588 [hep-ph].


\bibitem{Luo:2007bq}
  M.~Luo and G.~Zhu,
  arXiv:0704.3532 [hep-ph].

\bibitem{Chen:2007vv}
  C.~H.~Chen and C.~Q.~Geng,
  arXiv:0705.0689 [hep-ph].

\bibitem{Ding:2007bm}
  G.~J.~Ding and M.~L.~Yan,
  arXiv:0705.0794 [hep-ph].

\bibitem{Liao:2007bx}
  Y.~Liao,
  arXiv:0705.0837 [hep-ph].

\bibitem{Li:2007by}
  X.~Q.~Li and Z.~T.~Wei,
  arXiv:0705.1821 [hep-ph].


\bibitem{Aliev:2007qw}
  T.~M.~Aliev, A.~S.~Cornell and N.~Gaur,
  arXiv:0705.1326 [hep-ph].

\bibitem{Cheung:2007ap}
  K.~Cheung, W.~Y.~Keung and T.~C.~Yuan,
  arXiv:0706.3155 [hep-ph].



\bibitem{Yao:2006px}
  W.~M.~Yao {\it et al.}  [Particle Data Group],
  J.\ Phys.\ G {\bf 33}, 1 (2006).


\bibitem{Aubert:2006uy}
  B.~Aubert {\it et al.}  [BABAR Collaboration],
  Phys.\ Rev.\  D {\bf 75}, 031103 (2007)
  [arXiv:hep-ex/0607044].


\bibitem{Abbiendi:2001cs}
  G.~Abbiendi {\it et al.}  [OPAL Collaboration],
  Phys.\ Lett.\  B {\bf 519}, 23 (2001)
  [arXiv:hep-ex/0109011].

\end{thebibliography}
\end{document}